# A c=−2 boundary changing operator for the Abelian sandpile model

Philippe Ruelle
*Université catholique de Louvain*
*Institut de Physique Théorique*
*B–1348   Louvain-la-Neuve, Belgium*
(Dated: November 1, 2018)

We consider the unoriented two–dimensional Abelian sandpile model on the half–plane with open and closed boundary conditions. We show that the operator effecting the change from closed to open, or from open to closed, is a boundary primary field of weight $-1/8$, belonging to a $c = -2$ logarithmic conformal field theory.

## INTRODUCTION

The sandpile models are among the simplest and most studied non–equilibrium models showing criticality. It was introduced by Bak, Tang and Wiesenfeld [1] as a prototype of a class of dynamical models with generic criticality.

The model we consider here is the standard, unoriented two–dimensional Abelian sandpile model (ASM), which we first briefly recall. We refer to the recent reviews [2, 3] for a more complete account. The model is defined on a $L \times M$ square lattice. At each site $i$, there is a random variable $h_i$, which counts the number of grains of sand at $i$. A stable configuration is a set of $h_i \in \{1, 2, 3, 4\}$.

The discrete time evolution is defined as follows. First, one grain of sand is dropped on a random site of the current configuration $\mathcal{C}_t$, producing a new configuration $\mathcal{C}'_t$ which may not be stable. If $\mathcal{C}'_t$ is stable, we simply set $\mathcal{C}_{t+1} = \mathcal{C}'_t$. If $\mathcal{C}'_t$ is not stable (the new $h_i$ is equal to 5), $\mathcal{C}'_t$ relaxes to a stable configuration $\mathcal{C}_{t+1}$ by letting all unstable sites topple: a site with height $h_i \geq 5$ loses 4 grains of sand, of which each of its neighbours receives 1, something we can write in the form $h_j \to h_j - \Delta_{ij}$ for all sites $j$. $\Delta$ is called the toppling matrix, equal to the discrete Laplacian. Relaxation stops when no unstable site remains; the corresponding stable configuration is $\mathcal{C}_{t+1}$.

Two natural boundary conditions may be imposed along the boundaries: open and closed. A boundary site is open if it loses 4 grains in a toppling, his three (or two) neighbours receiving each 1 grain. It is closed if the number of grains it loses is exactly equal to the number of neighbours. Thus an open site is dissipative, whereas a closed site is conservative (like all bulk sites). In all cases, the toppling updating rule $h_j \to h_j - \Delta_{ij}$ applies, with $\Delta$ appropriately defined. The dynamics is well–defined provided that not all sites are conservative.

Dhar gave in [4] a first detailed analytical treatment of the model. Under very mild assumptions, he showed that on a finite lattice, there is a unique probability measure $P^*$ on the set of stable configurations, that is invariant under the dynamics. Moreover, $P^*$ is uniform on its support, formed by the so–called recurrent configurations. The partition function, defined as the number of recurrent configurations, is equal to $\det \Delta$, and goes like $(3.21)^{LM}$.

It is widely believed that at least some of the critical properties can be accounted for by a conformal field theory (CFT), despite the fact that the ASM show intrinsic non local features. The CFT picture has been assumed and used by various authors, but very few detailed, explicit comparisons have been made.

Our aim in this Letter is two–fold. First, we add strong support to the view that CFT is an appropriate description by comparing boundary CFT predictions with ASM calculations. Second, and because the relevant CFT is logarithmic with central charge $c = -2$, the ASM provides a concrete, lattice realization of such a non–unitary theory. In this respect, the cylinder partition functions computed below may shed some light on the structure of boundary states in logarithmic conformal theories, a topic of current interest [5].

## THE CONFORMAL FIELD THEORY

Various results have pointed to a relationship between the 2d ASM and the $q \to 0$ critical Potts model [6] or the $c = -2$ explicit Lagrangian realization [2],

$$S = \frac{1}{\pi} \int \partial\theta \, \bar{\partial}\bar{\theta} \,, \qquad (1)$$

where $\theta$ and $\bar{\theta}$ are scalar Grassmanian fields. The relevance of this field theory was confirmed in [7] where the unit height variables and other local cluster variables were indeed shown to go in the scaling limit to combinations of $\theta$'s and derivatives (the off–critical regime was also investigated in [7], where a relationship with the massive extension of (1) was established).

From the CFT point of view, $c = -2$ is special as it is the simplest example of a logarithmic CFT (and is necessarily non unitary) [8]. It has been discussed by many authors, from different points of view, and gives rise to many subtleties. For that matter, we refer the reader to the recent Tehran lecture notes by Flohr [9] and Gaberdiel [10], and to the references therein.

The free theory (1) has a $\mathcal{W}$–algebra generated by three dimension 3 fields and underlies a $c = -2$ conformal theory, which is rational with respect to the extended algebra [11, 12]. It contains six representations: $\mathcal{V}_{-1/8}$, $\mathcal{V}_0$, $\mathcal{V}_{3/8}$ and $\mathcal{V}_1$ are irreducible representations constructed out from ground states with conformal weight given by the subscript, and two reducible but indecomposable representations $\mathcal{R}_0$ and $\mathcal{R}_1$, each containing two ground states with zero conformal weight. All together they form a closed chiral fusion ring, but the characters of $\mathcal{V}_{-1/8}$, $\mathcal{V}_{3/8}$, $\mathcal{R}_0$ and $\mathcal{R}_1$ only are closed under modular transformations (and form a fusion subring). The chiral characters are computed in [13, 14]

$$\chi_{\mathcal{V}_0} = \frac{\theta_2}{4\eta} + \tfrac{1}{2}\eta^2, \qquad \chi_{\mathcal{V}_1} = \frac{\theta_2}{4\eta} - \tfrac{1}{2}\eta^2, \qquad (2)$$

$$\chi_{\mathcal{V}_{-1/8}} = \frac{\theta_3 + \theta_4}{2\eta}, \qquad \chi_{\mathcal{V}_{3/8}} = \frac{\theta_3 - \theta_4}{2\eta}, \qquad (3)$$

$$\chi_{\mathcal{R}_0} = \chi_{\mathcal{R}_1} = \frac{\theta_2}{\eta}, \qquad (4)$$

where the $\theta_i$ are the standard Jacobi theta functions, and $\eta$ is the Dedekind function [15].

## PARTITION FUNCTIONS

We want to compute $\mathbb{Z}_2$ orbifold partition functions for the free theory (1) on a cylinder, of perimeter $L$ and length $M$, and modulus $\tau = i\frac{L}{2M}$. Along the closed loop, we allow the two fields $\theta, \bar\theta$ to be either periodic or antiperiodic, and then sum over the two monodromies. On each boundary, we choose either open (Dirichlet) or closed (Neumann) boundary condition.

The calculations are fairly straightforward since the functional integral is just the determinant of the Laplacian, subjected to given monodromy and boundary conditions. For mixed boundary conditions, a simple calculation yields

$$\det -\Delta_{P/A,\text{mixed}} = \prod_{m \in \mathbb{Z},\, n \in \mathbb{Z}_+} 4\pi^2 [\tfrac{(m+\epsilon)^2}{L^2} + \tfrac{(n+1/2)^2}{4M^2}], (5)$$

with $\epsilon = 0, \tfrac{1}{2}$ for periodic resp. antiperiodic monodromy. These infinite products have been computed in [16]

$$\det -\Delta_{P,\text{mixed}} = \frac{\theta_4}{\eta}, \qquad \det -\Delta_{A,\text{mixed}} = \frac{\theta_3}{\eta}. \qquad (6)$$

The orbifold partition function, defined as half the sum of the two determinants, reduces to a single character of the triplet algebra at $c = -2$

$$Z_{\text{open,closed}} = \chi_{\mathcal{V}_{-1/8}}(q), \qquad q = e^{-\pi\frac{L}{M}}. \qquad (7)$$

When $L$ goes to infinity ($q \to 0$), the cylinder becomes an infinitely long strip with open and closed boundary condition on either side. The strip can be conformally mapped onto the upper half–plane, with the open and closed boundary conditions on either side of the origin, on the real axis. The change of boundary condition at the origin can then be seen as resulting from the action of the ground state of $\mathcal{V}_{-1/8}$ [17]. Thus the operator $\phi^{\text{op,cl}} = \phi^{\text{cl,op}}$ switching a boundary condition between open and closed is a primary field of weight $-\tfrac{1}{8}$. In the realization (1), this field is non local in $\theta, \bar\theta$.

Similarly one finds the partition functions for identical boundary conditions at both ends of the cylinder

$$Z_{\text{open,open}} = \chi_{\mathcal{V}_0}(q), \qquad Z_{\text{closed,closed}} = \chi_{\mathcal{R}_0}(q). \qquad (8)$$

$\mathcal{V}_0$ has the identity $\Omega$ as unique ground state, but $\mathcal{R}_0 \supset \mathcal{V}_0$ has two, namely $\Omega$ and its logarithmic partner $\omega$, proportional to $:\theta\bar\theta:$.

The above partition functions represent the finite size corrections to the large volume lattice partition functions, in the limit $L, M \to \infty$ with $\frac{L}{M} \to -2i\tau$, i.e. after the diverging bulk and boundary free energies have been subtracted. For what follows, it will be useful to know the leading term of the boundary free energies. For this purpose, it is more convenient to consider the partition function on a rectangle $L \times M$ (i.e. with no periodicity), with two edges open and two closed (the sides of length $M$ are closed say). The partition function is the determinant of the lattice Laplacian with appropriate boundary conditions (it can be seen as a lattice version of (1), or as the actual partition function of the 2d ASM). One finds

$$Z_{\{\substack{2\,\text{open}\\2\,\text{closed}}\}} = \prod_{m=0}^{L-1}\prod_{n=1}^{M}[4 - 2\cos\tfrac{m\pi}{L} - 2\cos\tfrac{n\pi}{M+1}]. \qquad (9)$$

The Euler–McLaurin formula can be used to evaluate the logarithm, and yields (see [18] for similar calculations)

$$\log Z_{\{\substack{2\,\text{open}\\2\,\text{closed}}\}} = L(M+1)\frac{4\text{G}}{\pi} - (L+M+1)\log(1+\sqrt{2})$$
$$+ \tfrac{1}{2}\log M + \log 2^{\tfrac{5}{4}}\,\eta(2\tau) + \ldots \qquad (10)$$

where the dots stand for terms that vanish in the limit $L, M \to \infty$, and where G $= 0.915965...$ is the Catalan constant.

The coefficient $\tfrac{4\text{G}}{\pi} = \log 3.21...$ of the most diverging term is the bulk free energy density, whereas the terms linear in $L$ and $M$ give the boundary free energy densities (remembering that the number of bulk sites is $(L-2)(M-2)$)

$$f_{\text{open}} = \frac{6\text{G}}{\pi} - \tfrac{1}{2}\log(1+\sqrt{2}),$$
$$f_{\text{closed}} = \frac{4\text{G}}{\pi} - \tfrac{1}{2}\log(1+\sqrt{2}). \qquad (11)$$

Therefore an open site has an excess of free energy equal to $\tfrac{2\text{G}}{\pi}$ with respect to a closed site. The entropy per site ($= e^f$) is then equal to 3.21 for a bulk site, 3.70 for an open boundary site and 2.07 for a closed boundary site.

## THE 2D ASM ON THE UPPER HALF–PLANE

We consider the ASM on the upper half–plane, $x \in \mathbb{Z}$, $y \geq 1$. Depending on the boundary condition, open or closed, the ASM is defined in terms of an infinite toppling matrix $\Delta_{\text{op}}$ or $\Delta_{\text{cl}}$. Away from the boundary, they coincide with the Laplacian on the plane; at a boundary site, their action is

$$(\Delta_{\text{op}} - 4) f(x,1) = (\Delta_{\text{cl}} - 3) f(x,1)$$
$$= -f(x,2) - f(x-1,1) - f(x+1,1). \quad (12)$$

We will make an extensive use of their inverse, $G_{\text{op}}$ and $G_{\text{cl}}$. By the method of images, one easily sees, for $i = (x_1, y_1)$, $j = (x_2, y_2)$, that ($x = x_2 - x_1$)

$$G_{\text{op}}(i;j) = G(x, y_2 - y_1) - G(x, y_2 + y_1), \quad (13)$$
$$G_{\text{cl}}(i;j) = G(x, y_2 - y_1) + G(x, y_2 + y_1 - 1), \quad (14)$$

in terms of the inverse $G$ of the Laplacian on the plane:

$$G(x,y) = \iint_0^{2\pi} \frac{d^2 k}{4\pi^2} \frac{e^{ixk_1 + iyk_2}}{4 - 2\cos k_1 - 2\cos k_2}. \quad (15)$$

The partition function of either ASM, defined as $Z_{\text{op}} = \det \Delta_{\text{op}}$ and $Z_{\text{cl}} = \det \Delta_{\text{cl}}$, clearly diverges. The same is true if one changes the boundary condition on a stretch of length $n$, by inserting $n$ closed sites in an open boundary, or $n$ open sites in a closed boundary. However one should expect that ratios be well–defined. Let us denote by $Z_{\text{op}}(n) = \det \Delta_{\text{op}}(n)$ the partition function of the ASM with open boundary condition everywhere on the boundary except on $n$ adjacent sites, which are closed, and by $Z_{\text{cl}}(n) = \det \Delta_{\text{cl}}(n)$ the partition function for the inverse situation.

We consider the ratios $Z_{\text{op}}(n)/Z_{\text{op}}$ and $Z_{\text{cl}}(n)/Z_{\text{cl}}$. They give the expectation value of the closing or opening of $n$ sites on the boundary, and so should correspond in the scaling regime to the 2–point function of the boundary changing operator discussed above. On the basis of CFT, we therefore expect that in the large $n$ limit,

$$\frac{Z_{\text{op}}(n)}{Z_{\text{op}}}, \frac{Z_{\text{cl}}(n)}{Z_{\text{cl}}} \longrightarrow \langle \phi(0) \phi(n) \rangle_{\text{CFT}} = A \, n^{\frac{1}{4}}, \quad (16)$$

for $A$ a normalization constant.

The explicit calculations described below confirm this limit, and the CFT picture behind it.

### Closed sites in open boundary

From (12), the operator $\Delta_{\text{op}}(n)$ only differs from $\Delta_{\text{op}}$ by the diagonal entries corresponding to the set $I$ of the $n$ sites which are being closed, $\Delta_{\text{op}}(n) = \Delta_{\text{op}} - B$, with $B_{i,j} = \delta_{i,j \in I}$. It follows that

$$\frac{Z_{\text{op}}(n)}{Z_{\text{op}}} = \det(\mathbb{I} - G_{\text{op}} B) = \det(\mathbb{I} - G_{\text{op}})_{i,j \in I} \quad (17)$$

is the determinant of a dimension $n$ matrix. Due to the horizontal translation invariance of $G_{\text{op}}$, the determinant has the Toeplitz form $\det(a_{i-j})$, with entries $a_m = \delta_{m,0} - G_{\text{op}}((0,1);(m,1))$ along diagonals. Using Eq.(13), a little of algebra gives $a_m$ as Fourier coefficients of the function (see Appendix A of [7] for material related to the lattice Green function)

$$\sigma(k) = \sqrt{(3 - \cos k)(1 - \cos k)} + \cos k - 1. \quad (18)$$

The Toeplitz determinant $\det(a_{i-j})$ may be computed by using a generalization of the Szegö limit theorem due to Widom [19]. Let $\sigma(k) = \tau(k)(2 - 2\cos k)^\alpha$, with $\alpha > -\frac{1}{2}$ and $\tau$ a single–valued, smooth, and nowhere vanishing nor divergent function on the unit circle, have the Fourier coefficients $s_m = \int_0^{2\pi} \frac{dk}{2\pi} e^{-ikm} \sigma(k)$. Then one has the asymptotic value of the Toeplitz determinant

$$\det(s_{i-j})_{1 \leq i,j \leq n} \sim E[\tau, \alpha] \, n^{\alpha^2} e^{n t_0}, \qquad n \gg 1. \quad (19)$$

The constant

$$E[\tau, \alpha] = e^{\sum_{m \geq 1} m t_m t_{-m}} [\tau(0)]^{-\alpha} \frac{G(\alpha + 1)^2}{G(2\alpha + 1)} \quad (20)$$

is explicit and given in terms of the Fourier coefficients $t_m$ of $\log \tau$ and the Barnes function $G$ [15].

Comparing with (18), Widom's theorem applies with $\alpha = \frac{1}{2}$ and $\tau(k) = \sqrt{\frac{3 - \cos k}{2}} - \sqrt{\frac{1 - \cos k}{2}}$. The Fourier coefficients of $\log \tau(k)$ are related to values of the inverse Laplacian on the plane (15)

$$t_m = \begin{cases} -\frac{2\mathrm{G}}{\pi} & \text{for } m = 0, \\ \frac{1}{2m}[G(m-1,0) - G(m+1,0)] & \text{for } m \neq 0. \end{cases} \quad (21)$$

The asymptotic ratio of the ASM partition functions is thus

$$\frac{Z_{\text{op}}(n)}{Z_{\text{op}}} = E[\tau; \tfrac{1}{2}] \, n^{\frac{1}{4}} \, e^{-\frac{2\mathrm{G}}{\pi} n}. \quad (22)$$

The exponential factor is due to the smaller free energy that a closed site has with respect to an open site, a difference equal to $\frac{2\mathrm{G}}{\pi}$, as we have seen earlier. This term must be subtracted before comparing with the corresponding CFT quantity. This leaves the expected power law with exponent $\frac{1}{4}$. The normalization is $A = E[\tau; \frac{1}{2}] = e^{\sum_{m \geq 1} m t_m t_{-m}} G(\frac{3}{2})^2 \sim 1.18894$, for which we have used the functional relation $G(z+1) = \Gamma(z) G(z)$ and $G(1) = 1$ [15].

### Open sites in closed boundary

In this case, the ratio of partition functions,

$$\frac{Z_{\text{cl}}(n)}{Z_{\text{cl}}} = \det(\mathbb{I} + G_{\text{cl}} B) = \det(\mathbb{I} + G_{\text{cl}})_{i,j \in I} \quad (23)$$

poses an immediate problem: all entries in this determinant are infinite, due to the divergent integral (15). From the ASM point of view, the divergence is a manifestation of non local effects and is caused by the breach we make in the boundary, which in effect frees an infinite number of configurations, which were until then forbidden and which now become recurrent. The divergence is present for $n=1$ and does not sharpen when $n$ increases.

This suggests to consider instead the ratio

$$\frac{Z_{\rm cl}(n)}{Z_{\rm cl}(1)} = \frac{1}{b_0}\,\det(b_{i-j})_{1 \leq i,j \leq n}, \qquad (24)$$

where the entries $b_m = \delta_{m,0} + G_{\rm cl}((0,1);(m,1))$ are the Fourier coefficients of $\sigma'(k) = \frac{1}{2} + \frac{1}{2}\sqrt{\frac{3-\cos k}{1-\cos k}}$. As explained above, only the ratio is well–defined, but the numerator and denominator may be regularized by replacing $\sigma'$ by

$$\sigma'_\alpha(k) = \tfrac{1}{2}(1-\cos k)^\alpha[\sqrt{1-\cos k} + \sqrt{3-\cos k}]. \qquad (25)$$

For $\alpha > -\frac{1}{2}$, all terms in (24) are finite, but both $b_0$ and the determinant develop a simple pole at $\alpha = -\frac{1}{2}$, so that the ratio is regular.

The singularity of $b_0$ is easily evaluated, yielding $b_0 = \frac{1}{2\pi(\alpha+1/2)} + $ finite. From (19), the pole of the determinant is all contained in the factor $G(2\alpha+1)^{-1} = \frac{\Gamma(2\alpha+1)}{G(2\alpha+2)} \sim \frac{1}{(2\alpha+1)}$. The coefficients $t'_m$ may be computed at $\alpha = -\frac{1}{2}$, where the function $\tau'(k) = \sqrt{\frac{1-\cos k}{2}} + \sqrt{\frac{3-\cos k}{2}} = \frac{1}{\tau(k)}$ is the inverse of the function encountered in the case "closed in open", so that the coefficients $t'_m = -t_m$ are minus those given in (21). Altogether one obtains

$$\frac{Z_{\rm cl}(n)}{Z_{\rm cl}(1)} = 2\pi {\rm Res}(E[\tau';\alpha];\alpha=-\tfrac{1}{2})\, n^{\frac{1}{4}}\, e^{\frac{2{\rm G}}{\pi}n}. \qquad (26)$$

By using (20), the prefactor is actually equal to $E[\tau;\frac{1}{2}]$, on account of $G(\frac{3}{2}) = \Gamma(\frac{1}{2})G(\frac{1}{2})$, and produces the same normalization constant $A$ for the boundary changing operator. As expected, the exponential factor now reflects the excess of boundary free energy of open boundary sites.

We should mention that the divergent factor $\frac{Z_{\rm cl}(1)}{Z_{\rm cl}}$ has a finite part, related to that of $G(0,0)$. Its value however depends on the shape of the rectangle, and thus has no well–defined infinite volume limit.

## THE FOUR–POINT FUNCTION

One may go further and consider the insertion of two stretches of closed sites in an open boundary (or vice–versa), and then compare it with the 4–point function of $\phi^{\rm cl,op}$. The conformal blocks of the latter are the complete elliptic integrals $K(x)$ and $K(1-x)$ [8]. The appropriate solution here is ($z_{ij} = z_i - z_j$, and $x = \frac{z_{12}z_{34}}{z_{13}z_{24}}$)

$$\frac{\langle\phi(1)\ldots\phi(4)\rangle}{A^2[z_{12}z_{34}]^{\frac{1}{4}}} = \frac{2}{\pi}(1-x)^{\frac{1}{4}}\,K(x). \qquad (27)$$

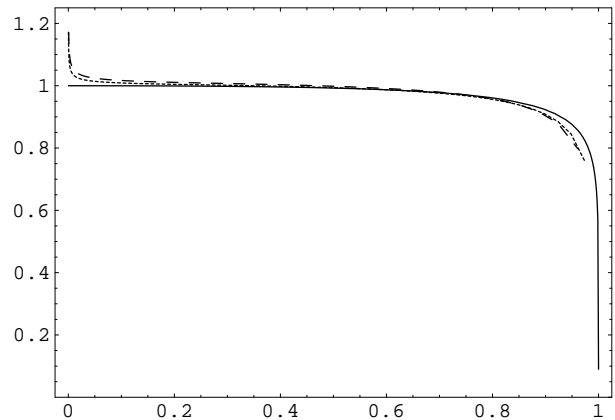

FIG. 1: Ratio of the ASM partition functions for two segments of closed sites ($[z_1,z_2]$ and $[z_3,z_4]$) in an open boundary, as function of the anharmonic ratio $x$. The solid curve is the CFT prediction, the other two are numerical.

This may be directly compared with a ratio of ASM partition functions. We consider an open boundary on which we close two sets of boundary sites, $I_1$ and $I_2$, ranging respectively over $[z_1,z_2]$ and $[z_3,z_4]$. Then the ratio

$$\frac{Z_{\rm op}(I_1,I_2)}{A^2[z_{12}z_{34}]^{\frac{1}{4}}}\,e^{\frac{2{\rm G}}{\pi}(|I_1|+|I_2|)} \sim \frac{Z_{\rm op}(I_1,I_2)\,Z_{\rm op}}{Z_{\rm op}(I_1)\,Z_{\rm op}(I_2)} \qquad (28)$$

is expected to converge to (27) in the scaling limit.

Exact ASM calculations are more difficult in this case, and we merely present in Fig.1 the results of (modest) numerical calculations. We have taken the two sets of closed sites to have equal length $n$ and to lie $N$ sites apart (i.e. $z_{21} = z_{43} = n$, $z_{32} = N$). We have fixed $2n+N$ to 200 and 300 (resp. long and short dashes), and in each case, we have let $n$ run so as to make $x = (\frac{n}{n+N})^2$ vary between 0 and 1. The numerical evaluation of (28) yields the dashed curves, while the solid line is the CFT result (27). The agreement is satisfactory in the region where the scaling regime is best approached.